\documentclass[aps,prd,nofootinbib,amsmath,amssymb,superscriptaddress,twocolumn]{revtex4} 
\usepackage{txfonts}
\usepackage{graphicx}
\usepackage{dcolumn}
\usepackage{bm}
\usepackage{amssymb}
\usepackage{latexsym}
\usepackage[colorlinks, linkcolor=blue, citecolor=blue, urlcolor=blue]{hyperref}

\newcommand{\be}{\begin{equation}}
\newcommand{\ee}{\end{equation}}
\newcommand{\bq}{\begin{eqnarray}}
\newcommand{\eq}{\end{eqnarray}}

\begin{document}

\title{Dark Energy and Fate of the Universe}

\author{Xiao-Dong Li}
\affiliation{Department of Modern Physics, University of Science and
Technology of China, Hefei 230026, China} \affiliation{Institute of
Theoretical Physics, Chinese Academy of Sciences, Beijing 100190,
China}
\author{Shuang Wang}
\affiliation{Department of Modern Physics, University of Science and
Technology of China, Hefei 230026, China} \affiliation{Institute of
Theoretical Physics, Chinese Academy of Sciences, Beijing 100190,
China}
\author{Qing-Guo Huang}
\affiliation{Institute of Theoretical Physics, Chinese Academy of
Sciences, Beijing 100190, China} \affiliation{Key Laboratory of
Frontiers in Theoretical Physics, Chinese Academy of Sciences,
Beijing 100190, China}
\author{Xin Zhang\footnote{Corresponding author}}
\email{zhangxin@mail.neu.edu.cn} \affiliation{Department of Physics,
College of Sciences, Northeastern University, Shenyang 110004,
China} \affiliation{Center for High Energy Physics, Peking
University, Beijing 100080, China}
\author{Miao Li}
\affiliation{Institute of Theoretical Physics, Chinese Academy of
Sciences, Beijing 100190, China} \affiliation{Key Laboratory of
Frontiers in Theoretical Physics, Chinese Academy of Sciences,
Beijing 100190, China}

\begin{abstract}
We explore the ultimate fate of the Universe by using a
divergence-free parametrization for dark energy $w(z)=w_0+w_a({\ln
(2+z)\over 1+z}-\ln2)$. Unlike the CPL parametrization, this
parametrization has well behaved, bounded behavior for both high
redshifts and negative redshifts, and thus can genuinely cover many
theoretical dark energy models. After constraining the parameter
space of this parametrization by using the current cosmological
observations, we find that, at the $95.4\%$ confidence level, our
Universe can still exist at least 16.7 Gyr before it ends in a big
rip. Moreover, for the phantom energy dominated Universe, we find
that a gravitationally bound system will be destroyed at a time $t
\simeq P\sqrt{2|1+3w(-1)|}/[6\pi |1+w(-1)|]$, where $P$ is the
period of a circular orbit around this system, before the big rip.
\end{abstract}

\pacs{95.36.+x, 98.80.Es, 98.80.-k}

\keywords{dark energy; dynamical dark energy; fate of the Universe}

\maketitle

There are two ultimate questions for human beings: ``where do we
come from?'' and ``where are we going?''. For a long time, they have
been topics of just religion and philosophy. But in the last three
decades, along with the rapid development of modern cosmology,
scientists have already obtained some important clues to these two
questions. To explain the origin of the Universe, cosmologists have
established a standard theoretical framework: Inflation + Hot Big
Bang. To foresee the destiny of the Universe, people have realized
that the key point is to understand the nature of dark
energy~\cite{DE}.

The fact we are facing is the absence of a consensus theory for dark
energy, though {\it ad hoc} models of dark energy based upon clever
ideas are not rare. In the absence of theoretical guidance, the
equation-of-state parameter (EOS) of dark energy, $w=p_{\rm
de}/\rho_{\rm de}$, provides a useful phenomenological description.
If cosmological observations could determine $w$ in a precise way,
then the underlying physics of dark energy would be successfully
revealed. However, in fact, extracting the information of $w$ from
the observational data is extremely difficult. Usually,
parametrization of dark energy is inevitable. A widely used approach
is the binned parametrization~\cite{Bin1}. The key idea is to divide
the redshift range into several bins and set $w$ as constant in each
bin. Since the current observational data at high redshifts (i.e.,
$z > 1$) are very rare, for the binned parametrization, only two
parameters of EOS $w$ can be constrained well~\cite{Bin2}. So in the
literature, a more popular approach is to assume a specific ansatz
for $w$.

Among all the ansatz forms of $w$, the Chevallier-Polarski-Linder
(CPL) parametrization~\cite{CP,Linder} is the most popular one. It
has a simple form,
\begin{equation}\label{eq1}
w(z)=w_0+w_a\frac{z}{1+z},
\end{equation}
where $z$ is the redshift, $w_0$ is the present-day value of the
EOS, and $w_a$ characterizes its dynamics. However, as pointed out
in our previous study~\cite{MZ1}, the CPL description will lead to
unrealistic behavior in the future evolution, i.e., $|w(z)|$ grows
rapidly and finally encounters divergence as $z$ approaches
$-1$~\cite{MZ1}. In order to keep the advantage of the CPL
parametrization, and avoid its drawback at the same time, we believe
that a divergence-free parameterization is necessary. In
Ref.~\cite{MZ1}, Ma and Zhang (MZ, hereafter, for convenience)
proposed the following hybrid form of logarithm and CPL
parametrizations:
\begin{equation}\label{eq2}
w(z)=w_0+w_a\left(\frac{\ln(2+z)}{1+z}-\ln2\right).
\end{equation}
This new parametrization has well behaved, bounded behavior for both
high redshifts and negative redshifts. In particular, for the
limiting case, $z\rightarrow -1$, a finite value for EOS can be
obtained, $w(-1)= w_0+w_a(1-\ln2)$.

According to the CPL description, the destiny of the Universe is
totally decided by the sign of $w_a$: if $w_a<0$, then
$w(-1)\rightarrow +\infty$, and so in the future the Universe will
again become matter dominated and return to decelerated expansion;
if $w_a>0$, then $w(-1)\rightarrow -\infty$, and so the Universe
will eventually encounter the ``big rip'' singularity. So, as we
have seen, the CPL description is unrealistic for predicting the
future of the Universe: the sign of $w_a$ solely determines the
final fate of the Universe that is vastly different for $w_a<0$ and
$w_a>0$. The same problem exists in other future-divergent
parametrizations, such as $w(z)=w_0+b\ln (1+z)$, $w(z)=w_0+w_a
z/(1+z)^2$, etc. For the MZ description, the future evolution of the
Universe is well-behaved \cite{MZ1,MZ2}, thus the current
constraints on the parameters might provide the important clue to
the ultimate fate of the Universe.

The theme of the fate of the Universe is rather profound and is not
testable. Nevertheless, the question of ``where are we going'' is so
attractive that we would like to make the inquiry. Our real purpose
is to highlight the importance of the detection of the dynamics of
dark energy. The future of the Universe might become conjecturable
if the dynamics of dark energy is identified by the data. Currently,
the observational data are not accurate enough to exclude or confirm
the cosmological constant, however, we still could infer how far we
are from a cosmic doomsday, in the worst case, from the current
data. In this study, we will try to discuss the topic of the destiny
of the Universe by analyzing the current data. It should be stressed
that our discussions depend on specific parametrizations of dark
energy. So, we do not claim that we are able to make robust
prediction for the future of the Universe. Instead we only use some
parametrizations to speculate about the future, basing on the
existing observational fact. We also assume that there will not be a
sudden change for the property of dark energy in the future.

Our focus will be on the MZ description since it is a
divergence-free parametrization. Of course, we will still discuss
the CPL description as a comparison. First, we will constrain the
cosmological models by using the current data. Next, we will discuss
the implications---from analyzing the data describing the past
history of the Universe---for the future of the Universe. The data
we used include the type Ia supernova (SN) data from the 3-yr SNLS
(SNLS3) observations~\cite{SNLS3}, the ``WMAP distance prior'' data
from the 7-yr WMAP (WMAP7) observations~\cite{WMAP7}, the baryon
acoustic oscillation (BAO) data from the SDSS Data Release 7 (SDSS
DR7)~\cite{SDSSDR7}, and the latest Hubble constant measurement from
the Hubble Space Telescope (HST)~\cite{H0}. We perform a $\chi^2$
analysis on the MZ and CPL models by using a Markov Chain Monte
Carlo technique~\cite{MCMC}.

\begin{figure}
\begin{center}
\includegraphics[width=8cm]{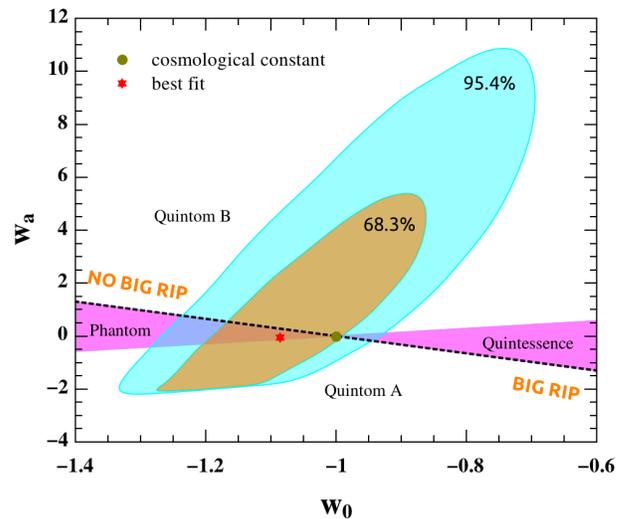}
\end{center}
\caption{\label{fig1} (color online). Probability contours at the
$68.3\%$ and $95.4\%$ CL in the $w_0$--$w_a$ plane, for the MZ
parametrization. The round point denotes the cosmological constant,
and the star symbol denotes the best-fit point of the MZ model. The
dashed line represents a divide: the region above this line
corresponds to a Universe without a big rip, and the region below
this line corresponds to a Universe with a big rip. According to the
properties of dark energy, the $w_0$--$w_a$ plane is divided into
four parts: quintessence, phantom, quintom A (with big rip), and
quintom B (without big rip). The quintessence is disfavored (at
1$\sigma$ CL) by the current observations, while both the
cosmological constant, the phantom, and the quintom are consistent
with the current observational data.}
\end{figure}

According to the joint data analysis, in Fig.~\ref{fig1} we plot the
probability contours at $68.3\%$ and $95.4\% $ confidence levels
(CL) in $w_0$--$w_a$ plane, for the MZ parametrization. For $95.4\%$
CL, the values of the model parameters are $\Omega_{\rm m0} =
0.2641^{+0.0423}_{-0.0342}$, $w_0 = -1.0862^{+0.4689}_{-0.2993}$,
$w_a=-0.0567^{+13.8399}_{-2.2345}$, and $h =
0.7235^{+0.0436}_{-0.0420}$, giving $\chi^2_{\rm min} = 423.444$.
Moreover, according to the properties of dark energy, we divide the
$w_0-w_a$ plane into four parts: quintessence, phantom, quintom A
(with big rip), and quintom B (without big rip). It is seen that the
quintessence is almost disfavored (at 1$\sigma$ CL) by the current
data. Although the best-fit point of the MZ parametrization
corresponds to a phantom, both the cosmological constant, the
phantom, and the quintom are consistent with the current
observational data at 1$\sigma$ CL. Therefore, the current
observational data are still too limited to indicate the properties
of dark energy.

To make a comparison, we also constrain the CPL parametrization by
using the same observational data, and find that the corresponding
model parameters are $\Omega_{\rm m0} = 0.2646^{+0.0417}_{-0.0340}$,
$w_0=-1.0665^{+0.5169}_{-0.4131}$, $w_a =
-0.0911^{+1.5033}_{-3.4714}$, and $h=0.7242^{+0.0410}_{-0.0420}$,
giving $\chi^2_{\rm min} = 423.432$. Although the CPL
parametrization can also fit the current data well, its EOS will
finally encounter divergence as $z$ approaches $-1$. So we do not
think that the CPL parametrization is a realistic description for
the future evolution of the Universe. Notwithstanding, we will still
discuss the big rip feature of the Universe using this description.

If in the future the EOS of dark energy $-1\leq w<-1/3$, then the
fate of the Universe is obvious: the expansion continues forever;
though galaxies disappear beyond the horizon and the Universe
becomes increasingly dark, structures that are currently
gravitationally bound still remain unaffected. This possibility
exists in our fitting result, but this case is too tedious to be
attractive, and so we have nothing to discuss for this case. What we
are really interested in is the existence of the possibility of the
``cosmic doomsday''. We want to infer, from the current data, how
the worst situation would happen in the future of the Universe. If
the doomsday exists, how far are we from it? Before the big rip,
what time would the gravitationally bound systems be torn apart?

Adopting the MZ description, we can find out the parameter space
where the big rip would happen. In Fig.~\ref{fig1}, we have denoted
the region with a big rip that is below the dashed line. A numerical
calculation will easily tell us the time of the doomsday: $t_{\rm
BR}-t_0=103.5$ Gyr for the best-fit result, and $t_{\rm
BR}-t_0=16.7$ Gyr for the $95.4\%$ CL lower limit, where $t_{\rm
BR}$ denotes the time of big rip (when the scale factor blows up).
In other words, for the worst case (2$\sigma$ CL lower limit), the
time remaining before the Universe ends in a big rip is 16.7 Gyr. As
a comparison, we also consider the case of CPL description, though
it is unrealistic for the future evolution of the Universe, as
discussed above. For the CPL description, we find that $t_{\rm
BR}-t_0=9.6$ Gyr for the $95.4\%$ CL lower limit.

Bound objects in the Universe, such as stars, globular clusters,
galaxies, and galaxy clusters, are stabilized since they have
detached from the Hubble flow, and so their internal dynamics are
independent of the cosmic expansion. However, if in the future
$w<-1$, the density of dark energy will grow so that eventually the
internal dynamics of bound objects will be influenced by dark
energy. Once the density of dark energy exceeds that of any object,
the repulsive gravity of phantom energy overcomes the forces holding
the object together, and the object is torn apart.

\begin{figure}
\begin{center}
\includegraphics[width=8cm]{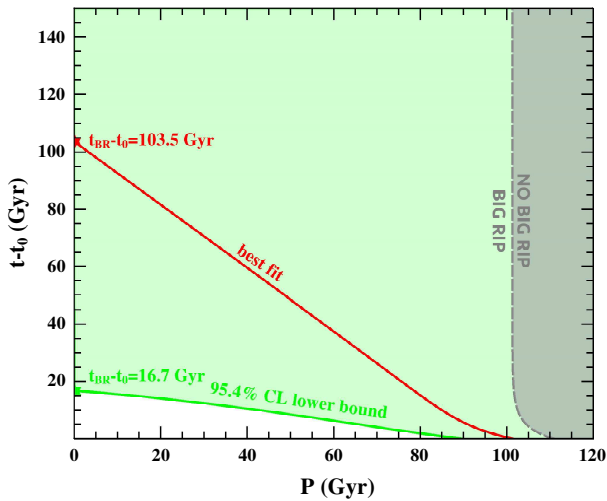}
\end{center}
\caption{\label{fig2} (color online). The relation between the
period of a gravitationally bound system $P$ and the rip time
$t-t_0$. The red line and the green line are plotted by using the
best-fit result and the $95.4\%$ CL lower limit of the MZ
parametrization, respectively. The dashed gray line is plotted by
using the cosmological constant model; the region on the left of
this line corresponds to a phantom dominated Universe, and the
region on the right of this line corresponds to a quintessence
dominated Universe.}
\end{figure}

For a gravitationally bound system with mass $M$ and radius $R$, the
period of a circular orbit around this system at radius $R$ is $P =
2\pi (R^{3}/GM)^{1/2}$, where $G$ is the Newton's constant.
As pointed out by Caldwell, Kamionkowski and Weinberg~\cite{bigrip},
this system will become unbound roughly when $-(4\pi / 3)(\rho_{\rm
de}+3p_{\rm de})R^3 \simeq M$. So one can determine the
corresponding redshift $z_{\rm tear}$ when this system is destroyed,
by solving the equation $[1+3w(z_{\rm tear})]f(z_{\rm tear})=-{8
\pi^2}/({\Omega_{\rm m0}H^{2}_{0}P^2})$, where
$f(z)=\exp[3\int_0^zdz'(1+w(z'))/(1+z')]$ characterizing the
dynamics of dark energy. According to the fitting results of the MZ
parametrization, we plot the relation between the characteristic
time scale of a gravitationally bound system $P$ and $t-t_0$ in
Fig.~\ref{fig2}. In fact, we are interested in the time interval
between the big rip and the event that a specific structure is
destroyed by phantom energy, which can be calculated by the integral
$t_{\rm BR}-t_{\rm tear}=\int^{z_{\rm
tear}}_{-1}[(1+z)H(z)]^{-1}{dz}$. In Ref.~\cite{bigrip}, Caldwell,
Kamionkowski and Weinberg have tackled this issue under the
framework of the constant $w$ model, and found a simple analytical
formula $t_{\rm BR}-t_{\rm tear} \simeq P\sqrt{2|1+3w|}/[6\pi
|1+w|]$, where $w={\rm constant}$. Now, the question is how to
handle the case of the dynamical dark energy (in the case of a
future phantom). We take the MZ parametrization as an example. In
this case, the function characterizing the dynamics of dark energy
has a simple approximate form, $f(z) \simeq
2^{6w_{a}}e^{-3w_{a}}(1+z)^{3(1+w(-1))}$, where
$w(-1)=w_0+w_a(1-\ln2)$. Then, we can easily derive a simple
analytical formula,
\begin{equation}\label{eq10}
t_{\rm BR}-t_{\rm tear} \simeq P \frac{\sqrt{2|1+3w(-1)|}}{6\pi
|1+w(-1)|}.
\end{equation}
Interestingly, this formula is very similar to the formula of
Ref.~\cite{bigrip}, except that the constant $w$ is replaced with
the value of $w(-1)$. This time interval is independent of $H_0$ and
$\Omega_{\rm m0}$, as expected. Numerical calculation shows that
this result is very precise. So we have shown that the formula of
Ref.~\cite{bigrip} can be extended to the case of a dynamical dark
energy (behaving as a phantom in the future).

\begin{figure}
\begin{center}
\includegraphics[width=8cm]{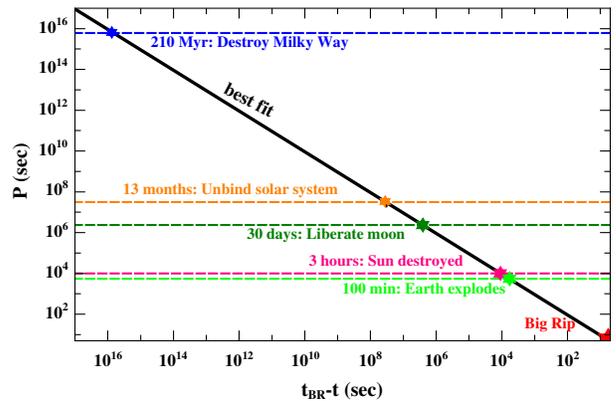}
\end{center}
\caption{\label{fig3} (color online). The relation between the
period of gravitationally bound system $P$ and the time interval
$t_{\rm BR}-t_{\rm tear}$. The solid black line is plotted by using
the best-fit result of the MZ parametrization. The dashed horizontal
lines are plotted by using the periods of various gravitationally
bound objects, including the Milky Way, the solar system, the
Earth-moon system, the Sun and the Earth.}
\end{figure}

In the following, let us speculate on a series of possible
consequences before the end of time. We shall describe when the
gravitationally bound systems (such as the Milky Way, the solar
system, the Earth-moon system, the Sun, the Earth, and so on) would
be destroyed. Utilizing the best-fit results of the MZ
parametrization, we plot the relation between the characteristic
time scale of gravitationally bound system $P$ and the time interval
$t_{\rm BR}-t_{\rm tear}$ in Fig.~\ref{fig3}. As is seen in this
figure, the Milky Way will be destroyed 210 Myr before the big rip;
13 months before the doomsday, the Earth will be ripped from the
Sun; 30 days before the doomsday, the moon will be ripped from the
Earth; the Sun will be destroyed three hours before the end of time;
and 100 min before the end, the Earth will explode.

\begin{table}
\caption{\label{tab1} The future of the Universe, where the $95.4\%$
CL lower limit of the MZ parametrization is used.}
\begin{ruledtabular}
\begin{tabular}{ll}
Time & Event\\
\hline
$t_0$ & Today \\
$t_{\rm BR} - 32.9$~Myr & Destroy Milky Way\\
$t_{\rm BR} - 2$~months & Unbind Solar System\\
$t_{\rm BR} - 5$~days & Strip Moon\\
$t_{\rm BR} - 28$~minutes & Sun Destroyed\\
$t_{\rm BR} - 16$~minutes & Earth Explodes\\
$t_{\rm BR} - 3 \times 10^{-17}$~s & Dissociate Atoms\\
$t_{\rm BR}-t_0 = 16.7$~Gyr &  Big Rip\\
\end{tabular}
\end{ruledtabular}
\end{table}

Actually, we are more interested in the worst situation. In
Table~\ref{tab1}, we list the corresponding events in the worst case
concerning the future of the Universe, where the results of the
$95.4\%$ CL lower limit are used. In this case, the Milky Way will
be destroyed 32.9 Myr before the big rip; 2 months before the
doomsday, the Earth will be ripped from the Sun; 5 days before the
doomsday, the moon will be ripped from the Earth; the Sun will be
destroyed 28 min before the end of time; and 16 min before the end,
the Earth will explode. Even microscopic objects cannot escape from
the rip. For example, the hydrogen atom will be torn apart $3 \times
10^{-17}$ s before the ultimate singularity.

In summary, we explored the ultimate fate of the Universe by using
the MZ parametrization. Unlike the CPL parametrization, the MZ
parametrization has well behaved, bounded behavior for both high
redshifts and negative redshifts, and thus can genuinely cover many
theoretical dark energy models. After constraining the parameter
space of this parametrization by using the current cosmological
observations, SN (SNLS3) + CMB (WMAP7) + BAO (SDSS DR7) + $H_0$
(HST), we found that, at the $95.4\%$ CL, our Universe can still
exist at least 16.7 Gyr before it ends in a big rip. Moreover, we
also discussed when a gravitationally bound system will be
destroyed, if our Universe is dominated by a phantom in the future.
It is found that a gravitationally bound system will be ripped apart
at a time $t \simeq P\sqrt{2|1+3w(-1)|}/[6\pi |1+w(-1)|]$ before the
big rip. This means that the conclusion of Ref.~\cite{bigrip} can be
extended to the case of dynamical dark energy.

For the CPL parametrization that is a future-divergent description,
we showed that the fate of the Universe is solely decided by the
sign of $w_a$. Obviously, this is unnatural. Nonetheless, we also
tested this possibility. For the CPL description, we found that, at
the $95.4\%$ lower limit, the big rip is from today by 9.6 Gyr; all
the bound systems in the Universe will be torn apart almost at the
same time---the big rip.

Of course, one might criticize that any prediction of the future of
the Universe is not testable and cannot be truly model-independent.
Nonetheless, we feel that, since the ultimate fate of the Universe
is closely interrelated to the nature of dark energy, it is fairly
natural to infer the future of the Universe from the current
detection of the dynamical property of dark energy. The question of
``where are we going'' is an eternal theme for human beings, so we
should have courage to explore this theme. Our attempt implies that
a rational parametrization for dynamical dark energy is important.
Whether a parametrization is divergence-free may or may not be an
important principle, but we have shown that a future divergent
description will lead to a strange future for the Universe, while a
divergence-free description can be used to foresee the cosmic
destiny in a rational way.

In addition to the MZ parametrization, one may also construct some
other parametrization forms to avoid the future divergence problem.
We only took the MZ form as a typical example. We also tested the
3-parameter form $w(z)=w_0+w_a z/(n+z)$. However, we found that $n$
cannot be constrained well by the data. To our current knowledge,
the MZ parametrization is a fairly good ansatz form to explore the
dynamics of dark energy as well as the fate of the Universe. Of
course, we expect to find out a better parametrization for probing
the dynamics of dark energy.


\begin{acknowledgments}
QGH is supported by the project of Knowledge Innovation Program of
Chinese Academy of Sciences and the Natural Science Foundation of
China (NSFC) under Grant No.~11105053. XZ is supported by the NSFC
under Grant Nos.~10705041, 10975032 and 11175042, and by the
National Ministry of Education of China under Grant
Nos.~NCET-09-0276 and N100505001. ML is supported by the NSFC under
Grant Nos.~10535060, 10975172 and 10821504, and by the 973 program
(Grant No.~2007CB815401) of the Ministry of Science and Technology
of China.
\end{acknowledgments}


\end{document}